\documentclass[12pt,twoside]{article}
\usepackage[mathscr]{eucal}
\usepackage{amsmath,amsfonts,amssymb,amsthm,mathabx,empheq}
\usepackage{times}
\usepackage{pdfsync}
\usepackage{cite}
\usepackage{url}
\usepackage{hyperref}
\usepackage{tensor}
\usepackage{color}
\usepackage{collref}
\advance\textheight\topskip
\allowdisplaybreaks[1]

\newcommand\td{\text{d}}
\newcommand\cO{{\cal O}}
\newcommand{\p}{\partial}

\newcommand{\be}{\begin{equation}}
\newcommand{\ee}{\end{equation}}
\newcommand{\bea}{\begin{eqnarray}}
\newcommand{\eea}{\end{eqnarray}}

\def\bz{\bar z}

\newcommand{\skyp}{{\cal I}^+}
\newcommand{\skym}{{\cal I}^-}
\newcommand\RR{\ensuremath{\mathbb R}}

\newcommand*\xbar[1]{%
  \hbox{%
    \vbox{%
      \hrule height 0.5pt % The actual bar
      \kern0.3ex%         % Distance between bar and symbol
      \hbox{%
        \kern-0.0em%      % Shortening on the left side
        \ensuremath{#1}%
        \kern-0.0em%      % Shortening on the right side
      }%
    }%
  }%
}

\DeclareFontFamily{OT1}{rsfs}{} \DeclareFontShape{OT1}{rsfs}{m}{n}{
<-7> rsfs5 <7-10> rsfs7 <10-> rsfs10}{}
\DeclareMathAlphabet{\mycal}{OT1}{rsfs}{m}{n}

\begin{document}
\title{Note on electromagnetic memories}

\author{Pujian Mao}

\date{}

\def\mytitle{Note on electromagnetic memories}

\pagestyle{myheadings} \markboth{\textsc{\small P.~Mao}} {\textsc{\small log memory}}
\addtolength{\headsep}{4pt}

\begin{centering}

  \vspace{1cm}

  \textbf{\Large{\mytitle}}

  \vspace{1.5cm}

  {\large Pujian Mao}%$^{\clubsuit}$$^{\diamondsuit}$

\vspace{.5cm}

\vspace{.5cm}
\begin{minipage}{.9\textwidth}\small \it  \begin{center}
     Center for Joint Quantum Studies and Department of Physics,\\
     School of Science, Tianjin University, 135 Yaguan Road, Tianjin 300350, China
 \end{center}
\end{minipage}

\end{centering}

\vspace{1cm}

\begin{center}
\begin{minipage}{.9\textwidth}
  \textsc{Abstract}. In this note, we provide a tractable example of a polyhomogeneous solution space for electromagnetism at null infinity in four dimensions. The memory effect for electromagnetism is then derived from the polyhomogeneous solution space. We also comment on the connection between the electromagnetic memories and asymptotic symmetries.
 \end{minipage}
\end{center}
\thispagestyle{empty}

%\newpage

\section{Introduction}

Memory effects broadly exist in gauge and gravity theories. In gravity theories, the memory effects can be observed by geodesic deviation \cite{memory,Braginsky:1986ia,1987Natur,Christodoulou:1991cr,Wiseman:1991ss,Thorne:1992sdb,Frauendiener}, time delay \cite{Pasterski:2015tva,Mao:2018xcw,Flanagan:2018yzh}, and velocity kick \cite{Grishchuk:1989qa,Podolsky:2002sa,Podolsky:2010xh,Podolsky:2016mqg,Zhang:2017rno,Zhang:2017jma,Zhang:2018srn,Hamada:2018cjj,Compere:2018ylh}. When the theory is solved in series expansion in the inverse powers of a radial coordinate $r$, the geodesic deviation has subleading order terms in the expansion known as infinite towers of memory \cite{Compere:2019odm} (see also \cite{Mao:2020vgh,Mirzaiyan:2020axm}). In non-abelian gauge theory, the memory effect is observed by a net relative color rotation of a pair of nearby quarks \cite{Pate:2017vwa,Ball:2018prg} (see also \cite{Jokela:2019apz,Campoleoni:2019ptc,Jokela:2020ibz}). In electromagnetism, the memory effects can be observed by velocity kick \cite{Bieri:2013hqa} and position displacement \cite{Mao:2017axa} of a test charged particle. The realization in experimental detections of memory effect can be found, for instance, in \cite{Susskind:2015hpa,Lasky:2016knh,Nichols:2017rqr,Yang:2018ceq,Bachlechner:2019deb,Madler:2016ggp,Madler:2017umy,Madler:2018tkl,Madison:2020xhh}.

In the context of the triangle relations \cite{Strominger:2017zoo}, memory effects are mathematically equivalent to soft theorems. Considering soft theorems as factorization properties that scattering amplitudes must obey in a low-energy expansion, soft factors should be related to memories in the $\frac1r$ expansion \cite{Compere:2019odm,Hamada:2018vrw}. In general, loop corrections involve logarithms of the energy of the soft particle. This has been precisely verified in the classical soft theorem \cite{Laddha:2018myi,Sahoo:2018lxl,Sahoo:2020ryf}. It is natural to ask whether there is logarithmic memory effect.

Memory effects are mostly investigated by asymptotic analysis, namely the theories are studied in series expansion. Though the expansion of the fields are typically of integer powers of the inverse of the radial coordinate \cite{Bondi:1962px}, there is a more realistic class of expansions involving logarithms, i.e., the polyhomogeneous expansion \cite{Chrusciel:1993hx}. In this note, we study electromagnetic memories with a special emphasis on the logarithmic term in $r$. This model allows one to illustrate several aspects of the logarithmic memory in a simplified setting. We obtain a self-consistent polyhomogeneous solution space for electromagnetism in four dimensions. We provide the exact memory formulas up to the first order with $\ln r$. The logarithmic term only involves the logarithmic contribution from the local source. The interpretations of such type of terms in literature are not in agreement. It is called null memory (kick) in \cite{Bieri:2013hqa} while memory effects only account for a modification induced by a burst of radiation in \cite{Compere:2019odm}. Here we would refer to the logarithmic term in the memory formula as \emph{logarithmic effect}. Then we derive an infinite tower of electromagnetic memories at all integer orders of $\frac1r$ when turning off the logarithmic terms. The memories at integer orders of $\frac1r$ are related to the large gauge transformation. However the logarithmic effect is not a transition between two vacua of the gauge field. It can NOT be associated with large gauge transformation.

The organization of this note is as follows. In the next section, we derive the polyhomogeneous solution space of electromagnetism. In section \ref{memory}, we specify the electromagnetic memories, in particular the logarithmic effect. In section \ref{symmetry}, we comment on the relation of electromagnetic memory and large gauge transformation. We close with a discussion in the last section.

\section{The polyhomogeneous solution space}

The Minkowski space-time has two null boundaries, past null infinity $\skym$, and future null infinity $\skyp$. They are better appreciated in advanced or retarded coordinates respectively. We will concentrate on $\skyp$ in the present work, although everything can be similarly repeated on $\skym$. The retarded spherical coordinates are defined with the change of coordinates as follows:
\begin{equation}
\label{retard}
u'=t-\sqrt{x^ix_i}\,,\quad r'=\sqrt{x^ix_i}\,,\quad x^1+ix^2=\frac{2r'z'}{1+z'{\bz}'}\,,\quad x^3=r\,\frac{1-z'{\bz}'}{1+z'{\bz}'}\,.
\end{equation}
The line element of Minkowski space-time becomes
\begin{equation}
\label{sphere}
\td s^2=-\td {u'}^2-2\td u'\, \td r' + \frac{2 {r'}^2}{P^2_S}\td z'\td{\bz}'\,,\qquad P_S=\frac{ 1+z'{\bz}' }{\sqrt2}\,.
\end{equation}
In the retarded spherical coordinates, $\skyp$ is just the submanifold $r=\infty$ with topology $S^2\times\RR$. For computational simplicity, the celestial sphere at null infinity can be mapped to a 2d plane with the following change of coordinates \cite{Barnich:2016lyg,Compere:2016jwb,Ball:2018prg,Barnich:2021dta}
:
\be\label{plane}\begin{split}
&u=P_S\left[u' - \frac{z' {\bz}' {u'}^2}{2(r' + P_S \frac{u'}{\sqrt2})}\right]\,,\\
&r=\frac{1}{P_S} \left(r' + P_S \frac{u'}{\sqrt2}\right)\,,\\
&z=z' - \frac{P_S z' u'}{\sqrt2(r' + P_S \frac{u'}{\sqrt2})}\,.\end{split}
\ee
The line element of Minkowski space-time now is given by
\be\label{flat}
\td s^2=-2 \td u \td r + 2r^2 \td z \td \bz .
\ee
We will work out the polyhomogeneous solution space of electromagnetism with this line element. Since the equations are in a covariant way, the results in line element~\eqref{sphere} can be simply derived by the change of coordinates~\eqref{plane}.

As \cite{He:2014cra}, we choose the following gauge and asymptotic conditions for the Maxwell fields and the current that is coupled to the Maxwell fields
\bea
\label{asycond}
  A_r=0 \ ,\;\;A_u=\cO(r^{-1}) \ ,\;\;A_z=\cO(1) \ ;\qquad J_r=0 \ ,\;\;J_u=\cO(r^{-2}) \ ,\;\;J_z=\cO(r^{-2}) \ .
\eea
A conserved current derived from a global symmetry is naturally defined up to the equivalence $J^\mu\thicksim J^\mu+\nabla_\nu k^{[\mu\nu]}$. Hence it makes more sense to consider equivalence classes of currents $[J^\mu]$ \cite{Barnich:2001jy}. We have used this ambiguity to set the radial component of the current to zero. This choice is more natural to work with the gauge choice of the Maxwell fields in~\eqref{asycond}.

It is convenient to arrange all Maxwell's equations in Minkowski space-time~\eqref{flat} with a conserved source $J^\mu$ as follows \cite{Barnich:2015jua,Conde:2016csj}:
\begin{itemize}
\item One hypersurface equation: $\nabla_\mu F^{\mu u}=J^u$,
\item Two standard equations: $\nabla_\mu F^{\mu z}=J^z$ and $\nabla_\mu F^{\mu\bz}=J^{\bz}$,
\item The current conservation equation: $\nabla_\mu J^\mu=0$,
\item One supplementary equation: $\nabla_\mu F^{\mu r}=J^r$.
\end{itemize}
\noindent When the first three types of equations are satisfied, the electromagnetic Bianchi equation $\nabla_\nu[\nabla_\mu F^{\mu\nu} - J^\nu]=0$ reduces to $\p_r[\sqrt{-g}( \nabla_\mu F^{\mu r} - J^r)]=0$. This implies that we just need to solve $\nabla_\mu F^{\mu r}= J^r$ at order $\cO(r^{-2})$, and all the remaining orders will automatically vanish. Thus the last equation is called the supplementary equation.

From the current conservation equation we get
\begin{equation}
\label{current}
  J_u=\dfrac{J^0_u(u,z,\bz)}{r^2} - \dfrac{1}{r^2} \int^{+\infty}_r \td r'\left(\p_z J_{\bz} + \p_{\bz} J_z\right) \ ,
\end{equation}
where $J^0_u(u,z,\bz)$ is the integration constant in $r$. Next, by integrating the hypersurface equation, we obtain
\begin{equation}
\label{Au}
  A_u=\dfrac{A^0_u(u,z,\bz)}{r} + \int^{+\infty}_r \td r'\, \frac{1}{r'^2} \int^{+\infty}_{r'}\td r''\left(\p_z \p_{r''} A_{\bz} + \p_{\bz}\p_{r''} A_z\right) \ ,
\end{equation}
where $A^0_u(u,z,\bz)$ is the integration constant and the other integration constant is turned off by the asymptotic condition~\eqref{asycond}. Let us assume the following ansatz for the expansion of the gauge field
\begin{equation}
\label{Azzb}
	A_{z(\bz)}=A^0_{z(\bz)}(u,z,\bz) + \sum\limits_{m=1}^\infty\sum\limits_{n=0}^{m} \frac{A^{m n}_{z(\bz)}(u,z,\bz)(\ln r)^n}{r^m} \ ,
\end{equation}
and the current
\begin{equation}
\label{Jzzb}
	J_{z(\bz)}=\dfrac{J^0_{z(\bz)}(u,z,\bz)}{r^2} + \sum\limits_{m=1}^\infty\sum\limits_{n=0}^m \frac{J^{m n}_{z(\bz)}(u,z,\bz)(\ln r)^n}{r^{m+2}} \ .
\end{equation}
The appearance of logarithmic terms indicates that the solutions are not smooth \cite{Chrusciel:1993hx}. A generic polyhomogeneous expansion includes also $n>m$ terms \cite{Chrusciel:1993hx}. However, we take a subset of the whole expansion proposed in \cite{Barnich:2015jua}, for which we can derive a finite logarithmic memory effect. If the current is generated by a collection of charged particles, the logarithmic terms in the asymptotic expansion represent the effect of long range electromagnetic interactions between the charged particles \cite{Laddha:2018myi,Sahoo:2018lxl}.

By integrating \eqref{current} and \eqref{Au} in $r$, we find that $u$-components of the gauge field and the current are solved as
\be
J_u=\dfrac{J^0_u(u,z,\bz)}{r^2} + \sum\limits_{m=1}^\infty\sum\limits_{n=0}^{m} \frac{J^{m n}_{u}(u,z,\bz)(\ln r)^n}{r^{m+2}}
\ee
and
\be
A_u=\dfrac{A^0_u(u,z,\bz)}{r} + \sum\limits_{m=1}^\infty\sum\limits_{n=0}^{m} \frac{A^{m n}_{u}(u,z,\bz)(\ln r)^n}{r^{m+1}}
\ee
where $J^{m n}_{u}$ and $A^{m n}_{u}$ are completely determined by $J^{m n}_{z(\bz)}$ and $A^{m n}_{z(\bz)}$. In particular, for the first order after the integration constant, we obtain
\begin{align}
&J^{11}_u=0\ ,\\
&J^{10}_u= -\p_z J_{\bz}^0 - \p_{\bz} J_z^0\ ,\\
&A_u^{11}=-\frac12 \p_{\bz} A_z^{11} - \frac12 \p_{z} A_{\bz}^{11} \ , \label{Au11}\\
&A_u^{10} =-\frac12 \p_{\bz} A_z^{10} - \frac12 \p_{z} A_{\bz}^{10} - \frac14 \p_{\bz} A_z^{11}  - \frac14 \p_{z} A_{\bz}^{11} \ .\label{Au10}
\end{align}

The time evolution of the coefficients of $A^{m n}_{z(\bz)}(u,z,\bz)$ is controlled by the standard equations which reduces to
\be\label{evolution}
\p_u \p_r A_z=\frac12 \p_r \p_z A_u + \frac{1}{2r^2} \p_z \left(\p_{\bz} A_z - \p_z A_{\bz}\right)-\frac12 J_z\ .
\ee
Clearly the retarded time derivative of all the coefficients in the expansion of $A_z$ have been uniquely determined except the leading $A^0_z$. We will refer to $\partial_u A^0_z$ as the \textit{news} function which reflects the propagating degree of freedom of electromagnetism. Changing $z\rightleftharpoons\bz$ above gives another \textit{news} function $\partial_uA^0_{\bz}$. We list several orders
\begin{align}
&\p_u A_z^{11}=0\ ,\label{11}\\
&\p_u A_z^{10}=\frac12 \p_z A_u^0 + \frac12 \p_z \left(\p_z A_{\bz}^0 - \p_{\bz} A_z^0\right) + \frac12 J_z^0\ ,\label{10}\\
&\p_u A_z^{22}=0\ ,\label{22}\\
&\p_u A_z^{21}=-\frac12 \p_z\p_{\bz} A_z^{11} + \frac14 J_z^{11} \ ,\label{21}\\
&\p_u A_z^{20}=-\frac14 \p_z \p_{\bz} A_z^{11} - \frac12 \p_z\p_{\bz} A_z^{10} + \frac18 J_z^{11} + \frac14 J_z^{10} \ .\label{20}
\end{align}
By turning off the logarithmic terms and mapping to the celestial sphere case, this result recovers the ones in \cite{Conde:2016csj}. The first piece on the right hand side of \eqref{21} will lead to divergence of $A_z^{21}$ at $u\rightarrow\pm\infty$. Hence $A_{\bz}^{11}$ should be set to zero from global properties \cite{Winicour:2014ska}. Then \eqref{Au11}, \eqref{Au10} and \eqref{21} are reduced to
\begin{align}
&A_u^{11}=0 \ , \\
&A_u^{10} =-\frac12 \p_{\bz} A_z^{10} - \frac12 \p_{z} A_{\bz}^{10} \ ,\\
&\p_u A_z^{21}= \frac14 J_z^{11} \ .
\end{align}

Finally, the supplementary equation gives the time evolution of the integration constant $A^0_u(u,z,\bz)$ as
\begin{equation}
\label{supeq}
  \p_u A^0_u  = \p_u(\p_z A^0_{\bz} + \p_{\bz} A^0_z) + J^0_u \ .
\end{equation}
To summarize, we have shown that the solution in polyhomogeneous expansion to the Maxwell system in four-dimensional Minkowski space-time~\eqref{flat} with the prescribed asymptotics~\eqref{asycond} is completely determined in terms of the initial data $A^0_u(u_0,z,\bz)$, $A^{mn}_z(u_0,z,\bz)$, $A^{mn}_{\bz}(u_0,z,\bz)$ ($m\geq1$), the news functions $A^0_z(u,z,\bz)$, $A^0_{\bz}(u,z,\bz)$ and the current $J^0_u(u,z,\bz)$, $J^{mn}_z(u,z,\bz)$, $J^{mn}_{\bz}(u,z,\bz)$ ($m\geq0$).

\section{The electromagnetic memories}
\label{memory}

Following closely the definition in \cite{Bieri:2013hqa}, the kick memory is induced by the time ($u$) integration of the electric field $E_z=\p_u A_z - \p_z A_u$ and its complex conjugate $E_{\bz}$. The electric field $E_z$ can be derived from the solution space in the previous section as
\be
E_z=E_z^0 + \frac{ E_z^1}{r} + \frac{ E_z^{21}\ln r}{r^2} +  \cO(r^{-2}),
\ee
where
\begin{align}
&E_z^0 = \p_u A_z^0 \ ,\label{first}\\
&E_z^1 = \p_u A_z^{10} - \p_z A_u^0 \ ,\label{second}\\
&E_z^{21} = \p_u A_z^{21} - \p_z A_u^{11} \ . \label{log}
\end{align}
The first order \eqref{first} has been well studied in literature. It is the E mode memory effect for which we can set $A_{z(\bz)}^0=\p_{z(\bz)} \alpha(u,z,\bz)$. Then \eqref{supeq} yields
\be
\p_z \p_{\bz} \delta \alpha = \frac12 \delta A_u^0 + \frac12 \int^{+\infty}_{-\infty} J_u^0 \td u,
\ee
where $\delta$ denotes the difference between late and early retarded times. The first and second pieces on the right hand side induce the linear and null kick respectively.

After some massaging, the second order becomes
\be
E_z^1= - \p_u A_z^{10} + \p_z(\p_z A_{\bz}^0 - \p_{\bz} A_z^0) + J_z^0.
\ee
After the $u$-integration, the first piece on the right hand side will induce a linear kick at the subleading order while the remaining two pieces on the right hand side are the null part.

Then the first logarithmic term appears at the sub-subleading order, which can be deduced to
\be\label{logmemory}
E_z^{21}=\frac14 J_z^{11}.
\ee
The logarithmic term only has the null part induced by the local source term.

\section{Comment on the relation to asymptotic symmetries}
\label{symmetry}

The residual (large) gauge transformation that preserving the conditions \eqref{asycond} is generated by an arbitrary function $\epsilon(z,\bz)$ on the 2d plane of the null infinity \cite{He:2014cra}. The action of the asymptotic symmetries on the solution space is quite simple. Infinitesimally, the non-zero components are just
\be\label{transf}
\delta_\epsilon A_z^0=\p_z \epsilon\ , \quad \delta_\epsilon A_{\bz}^0=\p_{\bz} \epsilon\ .
\ee
Though the electric field $E_z$ is gauge invariant, the memory formula is defined by its $u$-integration. For instance, the leading kick memory is just $\delta A_{z(\bz)}^0$, namely the passage of electromagnetic radiation through a region induces a transition from one configuration of $A_{z(\bz)}^0$ (vacuum) to another. The two different vacua of $A_{z(\bz)}^0$ are also related by the large gauge transformation $A_{z(\bz)}^0\rightarrow A_{z(\bz)}^0 + \Theta(u) \delta_\epsilon A_{z(\bz)}^0$ which reveals the equivalence of the memory effect and the asymptotic symmetry for electromagnetism \cite{Pasterski:2015zua}.

For the subleading memories at integer orders of $\frac1r$, the vacuum transition is implicitly given by the time evolution equations \eqref{evolution}. For instance, the next-to-leading order \eqref{second} in the vacuum case $J_\mu=0$ can be reorganized as
\be
E_z^1=-\p_z \p_{\bz} A_z^0\ ,
\ee
where we have used equations \eqref{10} and \eqref{supeq} and the integration constant in $u$ is set to zero for the global consideration. The memory formula is reduced to $\int \td u A_{z(\bz)}^0$ and is related to gauge transformation $A_{z(\bz)}^0\rightarrow A_{z(\bz)}^0 + \delta(u) \delta_\epsilon A_{z(\bz)}^0$. This computation can be extended to any higher order in analogue with the gravity case \cite{Mao:2020vgh}. In the vacuum case, the electric field $E_z$ at higher order in the expansions in integer powers of $\frac1r$ is given by\footnote{Note that we have turned off the logarithmic terms for this computation.}
\be
E_z^m =\left[ \frac1m \p_z^2 - \frac12 (m-1)\right] A_z^{m-1} \ , \quad m\geq2\ .
\ee
Hence the memory formula at $m$th order is reduced to $\int \td u A_{z(\bz)}^{m-1}$. The time evolution of $A_z^{m-1}$ is completely determined by $A_z^0$ through the time evolution equations
\be
\begin{split}
&\p_u A_z^1 = \p_z^2 A_{\bz}^0 \ ,\\
&\p_u A_z^m =-\left[ \frac1m\p_z \p_{\bz} + \frac12 (m-1)\right] A_z^{m-1} \ , \quad m\geq2\ .\\
\end{split}
\ee
Ignoring the integration constants in $u$, one obtains
\be
\begin{split}
& A_z^1 = \p_z^2 \int \td u' A_{\bz}^0 \ ,\\
& A_z^{m+2} =(-)^{m+1} \prod\limits_{k=0}^m \left[ \big[\frac{1}{k+2} \p_z \p_{\bz} + \frac12 (k+1)\big]\int \td u_{k+1} \right]\p_z^2 \int \td u' A_{\bz}^0 \ .\\
\end{split}
\ee
So the memory formula at $m$th order is related to gauge transformation $A_{z(\bz)}^0\rightarrow A_{z(\bz)}^0 + \frac{\td^m \Theta(u)}{\td u^m} \delta_\epsilon A_{z(\bz)}^0$.

Since the logarithmic term~\eqref{logmemory} only involves the local source term, it can not be generated by electromagnetic radiation. Hence the logarithmic effect is not connected by vacuum configuration of the gauge theory and can not be created by the presence of a burst of radiation between two given points at null infinity.
Memory observable with the latter property is in one-to-one correspondence with particular residual (large) gauge transformations \cite{Compere:2019odm}. So the logarithmic effect~\eqref{logmemory} is not related to any large gauge transformation. Moreover the theory is linear, the fields at integer orders of $\frac1r$ will not arise in the evolution of the logarithmic fields according to~\eqref{evolution}. In other words, the news functions will not arise in the evolution of the logarithmic fields. Hence all logarithmic effects in the present theory can not be related to any large gauge transformation.

\section{Discussions}

In this note, we have shown a consistent solution space in series expansion with logarithmic terms for electromagnetism in four dimensions. This has been applied to derive the logarithmic effect of the electromagnetic memory. There are infinite towers of electromagnetic memories at the integer orders of $\frac1r$ which are related to large gauge transformation.

The solution space we have derived is a subset of the most generic polyhomogeneous expansion. More logarithmic terms can be included in the initial data in \eqref{Azzb} and \eqref{Jzzb}. However the powers of logarithmic term in the expansion should be finite. The time evolution equation \eqref{evolution} will set all logarithmic terms at order $\cO(r^{-1})$ to zero.

The logarithmic effect that we have derived is not related to asymptotic symmetries. Nevertheless it is still of interest to study this type of memory effect in the context of the triangle relations, e.g., to see if it has any connection to the logarithmic terms in the soft theorem \cite{Laddha:2018myi,Sahoo:2018lxl,Sahoo:2020ryf}.

\section*{Acknowledgments}
The author thanks Shixuan Zhao for the early collaboration on this project. This work is supported in part by the National Natural Science Foundation of China under Grants No. 11905156 and No. 11935009.

\bibliography{ref}

\end{document}